\documentclass[trackchanges,twocolumn]{aastex7}
\usepackage{subcaption}
\usepackage{booktabs}
\usepackage[T1]{fontenc}
\usepackage{lmodern}
\usepackage{hyperref}

\newcommand{\jfw}[1]{{#1}}

\begin{document}

\title{State-Dependent X-ray Variability in Cygnus X-1: A 12-Year NuSTAR Timing Study of Accretion Flow Geometry}

\author{Kshitij Duraphe}
\affiliation{Boston University, 725 Commonwealth Avenue, Boston, MA 02215, USA}
\email[show]{kshitijd@bu.edu}  

\author[orcid=0009-0002-6037-4613]{Kartik Mandar}
\affiliation{Raman Research Institute, Bangalore}
\affiliation{Indian Institute of Science Education and Research Bhopal}
\email{kartik4321mandar@gmail.com}  

\author[orcid=0009-0009-1218-7451]{Chooda Khanal}
\affiliation{Florida International University, Miami, FL 33199, USA}
\affiliation{Miami Dade College, Miami, FL 33176, USA}
\email[show]{ckhanal@fiu.edu}  

\author{Abha Pareek} 
\affiliation{Osmania University, Department of Astrophysics}
\email{abhapareek1206@gmail.com}

\author{Tejaswi Kondhiya}
\affiliation{Physical Research Laboratory, Ahmedabad}
\affiliation{UN Center for Space Science and Technology Education in Asia and Pacific (CSSTEAP)}
\email{tejaswi.physics11@gmail.com}

\author{V Sree Suswara}
\affiliation{Sardar Vallabhbhai National Institute of Technology, Surat}
\email{sreesuswara2020@gmail.com}

\author{Deeksha Dinesh}
\affiliation{Pondicherry University, Puducherry}
\email{deeksha3698@gmail.com}

\author{Vidyasagar Bhat}
\affiliation{School of Sciences, JAIN (Deemed-to-be University), Bengaluru}
\email{vidyasagarbhat106@gmail.com}

\author[orcid=0000-0002-0705-6619]{Gopal Bhatta}
\affiliation{Janusz Gil Institute of Astronomy, University of Zielona Góra,\\
ul. Szafrana 2, 65-516 Zielona Góra, Poland}
\email[show]{g.bhatta@ia.uz.zgora.pl}

\begin{abstract}

We present a comprehensive timing analysis of the black hole X-ray binary Cygnus X-1 using 26 NuSTAR observations spanning 2012-2024, providing the most detailed characterization to date of its accretion flow variability across spectral states. Our analysis reveals fundamental insights into the physics governing state transitions in stellar-mass black holes. \jfw{We characterize the energy-dependent manifestation of the well-known bimodal state distribution, showing that raw count rate bimodality is intrinsic to the Comptonized spectral component above $\sim$10 keV, while thermal emission below 8 keV remains unimodal.} Power spectral analysis uncovers state-dependent characteristic frequencies shifting from 0.050 Hz (hard) to 0.074 Hz (intermediate), with featureless red noise in soft states. These frequencies correspond to disk truncation radii evolving from $\sim$5.5 $R_g$ to $\sim$2 $R_g$, providing direct observational evidence for the inward progression of the accretion disk during state transitions. Frequency-dependent time lags evolve systematically from $\sim$50 ms hard lags at 0.1 Hz in hard states to near-zero in soft states, quantifying the collapse of the Comptonizing corona. Linear rms-flux relations persist across all states with parameters that precisely track the relative contributions of thermal versus non-thermal emission components. Most remarkably, we identify a failed state transition (observation 30302019006) exhibiting anticorrelated band behavior, suppressed variability ($F_{var}$ < 1.38\%), and apparent sub-ISCO truncation. This discovery challenges standard transition models and suggests new pathways for accretion flow evolution in wind-fed systems.
\end{abstract}



\section{Introduction}\label{sec:intro}

Microquasars are X-ray binary (XRB) systems consisting of a compact object, such as a stellar-mass black hole or a neutron star, that is accreting matter from a stellar companion. A defining characteristic of these systems is their ability to launch collimated, relativistic jets of plasma \citep{1999ARA&A..37..409M}. The discovery and subsequent study of microquasars have provided profound new insights into the physics of accretion and ejection in the vicinity of compact objects, establishing them as scaled-down analogues of active galactic nuclei (AGNs) and quasars \citep{1998Natur.392..673M}. These systems exhibit physical processes analogous to those in quasars, but on a significantly smaller scale. \cite{2005astro.ph..6008C}. This analogy is founded on the common architecture of a central accreting object, an accretion disk, and bipolar jets, with physical processes scaling in proportion to the mass of the central object. Given that the timescales of variability in microquasars are millions of times shorter than in their supermassive counterparts, they serve as unique laboratories for observing the dynamics of disk-jet coupling in real time.

The first object recognized to possess relativistic jets was SS 433 \citep{1984ARA&A..22..507M}. The microquasar class was formally established with the discovery of radio jets from 1E 1740.7-2942 \citep{1992Natur.358..215M}, and confirmed by the observation of apparent superluminal motion in GRS 1915+105 \jfw{\citep{1994Natur.371...46M}} and GRO J1655-40 \citep{1995Natur.375..464H,1996ApJ...466L..63J}, solidifying the physical connection to quasars.

For Black Hole X-Ray Binaries (BHXRBs), observational campaigns have revealed a strong coupling between the state of the accretion flow, diagnosed by its X-ray spectral and timing properties (see \cite{2006ARA&A..44...49R} for a review), and the nature of the jet. In the low/hard X-ray state, BHXRBs typically exhibit a steady, self-absorbed, compact radio jet with a characteristic flat or inverted radio spectrum \citep{2001MNRAS.322...31F,2001ApJ...554...43C}. As the source transitions to the high/soft state, which is dominated by thermal emission from the accretion disk, this compact jet is quenched \jfw{\citep{1999ApJ...519L.165F}}. The transition between these states is often accompanied by major, discrete ejection events that produce transient jets observed in sources like GRS 1915+105 \citep{2004MNRAS.355.1105F}.

\section{Cygnus X-1: The Archetypal Black Hole Microquasar}
Cygnus X-1 is the prototypical high-mass X-ray binary (HMXB) and was the first source for which the accretor was dynamically confirmed to be a black hole \citep{1972Natur.235...37W, 1972Natur.235..271B}. The system consists of a black hole accreting from the focused stellar wind of its O9.7Iab supergiant companion, HDE 226868. Recent measurements place the system at a distance of $2.22 \pm 0.18$ kpc, and calculate a black hole mass of $21.2 \pm 2.2$ M$_{\odot}$ \citep{2021Sci...371.1046M}. Extensive observational campaigns have established that the black hole is rapidly rotating. Early X-ray reflection measurements found $a^* \approx 0.97^{+0.014}_{-0.02}$ \citep{2012MNRAS.424..217F}, consistent with continuum-fitting results of $a^* > 0.983$ \citep{2014ApJ...790...29G}. More recent analyses yield $a^* > 0.9985$ (3$\sigma$) \citep{Zhao_2021}, while combined methods report $a^* = 0.87$--$0.90$ depending on inclination assumptions \citep{2024ApJ...967L...9Z}. The consensus from multiple approaches points towards $a_{*} > 0.98$ \citep{2014ApJ...790...29G, 2016A&A...589A..14D}.

As a persistent source, Cygnus X-1 has served as a touchstone for defining the canonical spectral states of accreting black holes, primarily the hard and soft states \citep{2001MNRAS.321..759C}. Unlike transient systems, however, it never enters a fully disk-thermal-dominant state; its soft state is more accurately described as a steep power-law state\citep{annurev:/content/journals/10.1146/annurev.astro.44.051905.092532}, and the source remains on the lower, hard-state branch of the hardness-intensity diagram \citep{Steiner2024AnIX}. In the hard state, the spectrum is dominated by a power-law component produced by inverse Compton scattering in a hot corona, which exhibits a high-energy cutoff around 100 keV \citep{Cangemi:2019uti}. This state is accompanied by a persistent, compact relativistic jet \citep{2006MNRAS.369..603F} and a non-thermal, polarized tail extending to MeV energies, likely synchrotron emission from the jet base \citep{gammacygx-1}.

The source's timing behavior is also state-dependent. The hard state exhibits strong, broadband variability whose power spectrum is well-modeled by a superposition of Lorentzian components \citep{2000A&A...357L..17P}. In the soft state, the variability drops to $\sim 10$-20\% root mean square (rms) and is characterized by red noise \citep{Cui1996XrayVO}. The linear relationship between rms variability and flux observed across timescales suggests that accretion rate fluctuations propagate through the flow \citep{2001MNRAS.323L..26U}.

X-ray observations have been instrumental in characterizing the physical properties of Cygnus X-1's accretion flow and corona. Various X-ray observatories, including \textit{INTEGRAL}\citep{cygx1_integral}, \textit{Suzaku}\citep{2015ApJ...808....9P}, and \textit{Chandra}\citep{2011ApJ...728...13N}, have provided detailed spectral and timing analyses that have shaped our understanding of this system. X-ray spectroscopy has enabled detailed modeling of reflection components originating from material near the black hole. These analyses constrain parameters such as black hole spin, inner disk radius, and coronal properties \citep{2018cosp...42E3408T, ZHU2024381}.

The Nuclear Spectroscopic Telescope Array (\textit{NuSTAR}) \citep{2013ApJ...770..103H}, launched in 2012, has provided unprecedented sensitivity in the hard X-ray band (3--79 keV), offering new insights into high-energy processes occurring in Cygnus X-1. Its focusing optics deliver significantly improved spatial and spectral resolution compared to previous hard X-ray instruments.

\textit{NuSTAR} observations have been used to test fundamental aspects of general relativity by analyzing reflection features in spectra to constrain black hole spin parameters \citep{2015ApJ...808....9P}. For example, joint \textit{NuSTAR} and \textit{Suzaku} observations constrained Cygnus X-1's spin parameter to values between 0.861 and 0.921 \jfw{\citep{2022ApJ...934....4K}} using advanced models like \texttt{kerrC}, which account for complex coronal geometries\jfw{; these values are somewhat lower than the consensus $a_* > 0.98$ discussed above, likely reflecting systematic differences in coronal geometry assumptions}. Additionally, \textit{NuSTAR} analyses favor extended wedge-shaped coronae over compact lamp-post configurations for explaining observed spectra. These findings challenge simplified models while advancing our understanding of accretion flows near black holes.

\section{\textit{NuSTAR} Observations and Data Reduction}

We processed all \textit{NuSTAR} observations of Cygnus X-1 with issue flag 0 using the \textit{NuSTAR} Data Analysis Software (\texttt{NuSTARDAS}) provided within the HEASoft\jfw{\footnote{\url{https://heasarc.gsfc.nasa.gov/lheasoft/}}} environment, along with the corresponding Calibration Database (CALDB) June 30, 2025 update.  A log of these data is given in Table \ref{tab:obslog}. For each Observation ID (OBSID), the raw Level 1 data were reprocessed using the \texttt{nupipeline} tool. This standard pipeline task performs instrument calibration, applies coordinate transformations, and generates cleaned, calibrated Level 2 event files for both Focal Plane Modules, FPMA and FPMB. 

During pipeline execution, standard screening criteria were applied to the event data to filter for good time intervals (GTIs). We selected science-grade events and filtered out periods of high particle background identified by the CsI anti-coincidence shield. This procedure ensures the use of high-quality data and maximizes the signal-to-noise ratio.

Following the initial processing, we extracted scientific products from the cleaned Level 2 event files using the \texttt{nuproducts} tool. For both FPMA and FPMB in each observation, source products were extracted from a circular region with a radius of $120''$ centered on the nominal J2000 coordinates of Cygnus X-1. The background was estimated from a nearby, source-free circular region of the same size located on the same detector chip. We applied a barycentric correction to adjust photon arrival times to the solar system barycenter; this was performed by setting the \texttt{barycorr=yes} parameter in \texttt{nuproducts}. This process generated the necessary source and background spectra, light curves, ancillary response files (ARFs), and redistribution matrix files (RMFs).

The final, background-subtracted light curves for each observation were then produced using the \texttt{lcmath} FTOOL. For a given OBSID, the source light curves from FPMA and FPMB were summed to create a combined source light curve. The corresponding background light curves were similarly summed. The final light curve was then created by subtracting the combined, area-scaled background from the combined source, resulting in a single, high signal-to-noise light curve for each observation, ready for timing analysis.

To finalize the light curves for analysis, we first extracted them at multiple time resolutions ($0.1$ s, $1$ s, $10$ s, $100$ s, and $300$ s) to assess the balance between temporal detail and statistical quality. For the primary timing analysis, we selected the $0.1$ s resolution. This choice maximizes sensitivity to the rapid, aperiodic X-ray variability that is a key characteristic of accreting black hole systems like Cygnus X-1. For all analyses except two, we utilize the $0.1$ s observations. To mitigate the influence of outliers, we filtered the selected light curves by removing data points with count rates below the 1st percentile or above the 99th percentile. This conservative clipping procedure is designed to remove the most extreme statistical fluctuations in the tails of the count rate distribution, which could otherwise disproportionately affect variance-based analyses. We found that, on average, 0.01\% of points were removed per observation.

\begin{deluxetable*}{ccccc}
\caption{Log of NuSTAR observations used in this work}
\label{tab:obslog}
\tablehead{
\colhead{Start Date} & \colhead{End Date} & \colhead{ObsID} & \colhead{Exposure} & \colhead{Start Time (UTC)} \\
\colhead{(YYYY-MM-DD)} & \colhead{(YYYY-MM-DD)} & & \colhead{(s)} & \colhead{(HH:MM:SS)}
}
\startdata
\jfw{2012-10-31} & \jfw{2012-10-31} & \jfw{30001011002} & \jfw{10442} & \jfw{08:11:11} \\
\jfw{2012-11-01} & \jfw{2012-11-01} & \jfw{10014001001} & \jfw{4184} & \jfw{00:16:07} \\
\jfw{2014-04-29} & \jfw{2014-04-30} & \jfw{30001011005} & \jfw{13507} & \jfw{22:11:08} \\
\jfw{2014-05-20} & \jfw{2014-05-20} & \jfw{30001011007} & \jfw{34372} & \jfw{05:51:04} \\
\jfw{2014-10-04} & \jfw{2014-10-04} & \jfw{30001011009} & \jfw{20355} & \jfw{17:36:05} \\
\jfw{2015-01-19} & \jfw{2015-01-19} & \jfw{30001011011} & \jfw{16824} & \jfw{00:36:08} \\
\jfw{2015-05-27} & \jfw{2015-05-27} & \jfw{30101022002} & \jfw{19861} & \jfw{17:06:08} \\
\jfw{2016-02-11} & \jfw{2016-02-11} & \jfw{90101020002} & \jfw{13472} & \jfw{10:51:10} \\
\jfw{2016-05-27} & \jfw{2016-05-28} & \jfw{30002150002} & \jfw{50077} & \jfw{22:41:05} \\
\jfw{2016-05-29} & \jfw{2016-05-30} & \jfw{30002150004} & \jfw{50902} & \jfw{21:26:12} \\
\jfw{2016-06-02} & \jfw{2016-06-03} & \jfw{30002150008} & \jfw{28717} & \jfw{22:01:12} \\
\jfw{2016-07-18} & \jfw{2016-07-18} & \jfw{30202032002} & \jfw{13312} & \jfw{15:31:06} \\
\jfw{2017-11-04} & \jfw{2017-11-04} & \jfw{30302019002} & \jfw{9400} & \jfw{17:16:13} \\
\jfw{2018-02-08} & \jfw{2018-02-08} & \jfw{30302019004} & \jfw{12708} & \jfw{20:01:06} \\
\jfw{2018-03-26} & \jfw{2018-03-26} & \jfw{30302019006} & \jfw{11024} & \jfw{17:46:10} \\
\jfw{2018-05-27} & \jfw{2018-05-27} & \jfw{30302019010} & \jfw{8214} & \jfw{08:31:12} \\
\jfw{2018-08-11} & \jfw{2018-08-11} & \jfw{30302019012} & \jfw{12119} & \jfw{03:01:09} \\
\jfw{2019-08-06} & \jfw{2019-08-06} & \jfw{80502335002} & \jfw{13436} & \jfw{09:01:09} \\
\jfw{2019-11-13} & \jfw{2019-11-13} & \jfw{80502335006} & \jfw{11914} & \jfw{10:06:05} \\
\jfw{2022-05-20} & \jfw{2022-05-20} & \jfw{30702017006} & \jfw{12367} & \jfw{14:06:08} \\
\jfw{2022-06-20} & \jfw{2022-06-20} & \jfw{90802013002} & \jfw{13178} & \jfw{12:01:09} \\
\jfw{2022-06-21} & \jfw{2022-06-21} & \jfw{90802013004} & \jfw{13781} & \jfw{10:26:06} \\
\jfw{2023-05-24} & \jfw{2023-05-24} & \jfw{80902318002} & \jfw{13507} & \jfw{18:31:06} \\
\jfw{2023-06-14} & \jfw{2023-06-14} & \jfw{80902318004} & \jfw{9326} & \jfw{00:26:12} \\
\jfw{2024-04-08} & \jfw{2024-04-08} & \jfw{30901039002} & \jfw{16879} & \jfw{18:16:07} \\
\jfw{2024-07-12} & \jfw{2024-07-12} & \jfw{91002320004} & \jfw{13718} & \jfw{14:56:06} \\
\enddata
\end{deluxetable*}

\section{Analysis and Results}

\subsection{Timing Analysis}

\subsubsection{Long-Term Spectral Evolution}

\begin{figure*}
    \centering
    \includegraphics[scale=0.39]{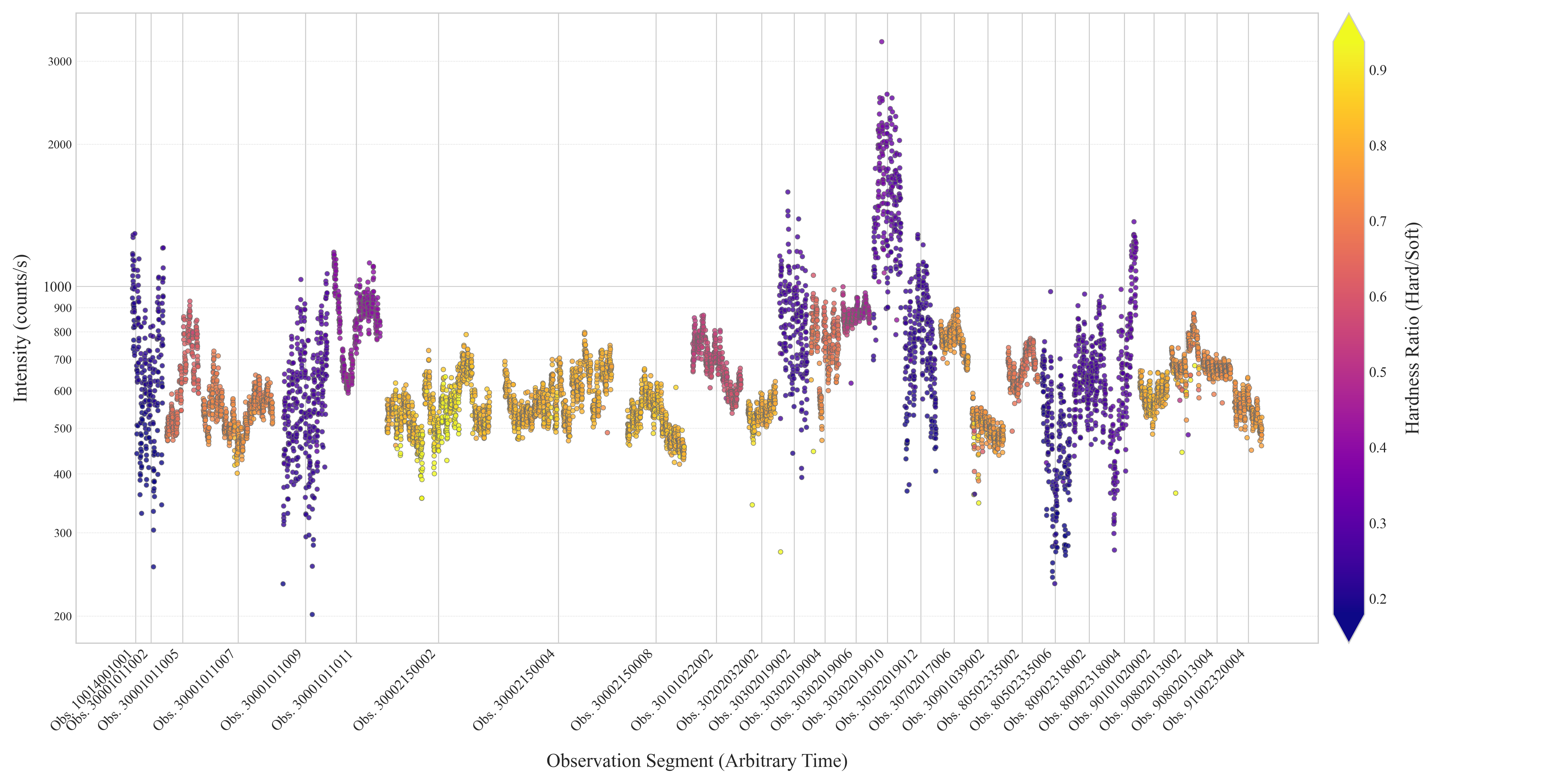}
    \caption{The long-term evolution of Cygnus X-1's intensity and spectral hardness. Each point represents a 100s time bin, with observation segments separated for visual clarity. The y-axis shows the total intensity in the 3-79 keV band, and the color of each point corresponds to the hardness ratio (HR), defined as the 8-79 keV rate divided by the 3-8 keV rate. Yellow/orange colors indicate a harder spectrum, while purple/blue colors indicate a softer spectrum, clearly showing the source transitioning between hard and soft states.}
    \label{fig:longtermlc}
\end{figure*}

We tracked the long-term X-ray evolution of Cygnus X-1 across the entire \textit{NuSTAR} observational campaign. Figure \ref{fig:longtermlc} presents the total intensity (3-79 keV) light curve, constructed from 100s time bins, with each point color-coded by its corresponding hardness ratio (HR). The x-axis represents the sequence of individual \textit{NuSTAR} observations, stitched together to visualize the source's behavior over time.

The source exhibits profound variability in both its flux and spectral shape, with multiple transitions between the canonical hard and soft spectral states clearly visible. The hard state is prominent throughout many observations (e.g., Obs. 30002150002), characterized by a relatively low intensity of approximately 400-800 counts/s and a high HR, indicated by yellow and orange points. Conversely, the soft state manifests as periods of high intensity (>1000 counts/s) and a low HR, indicated by purple and blue points. A particularly dramatic flare event is captured in Obs. 30302019010 and Obs. 30302019012, where the source intensity exceeds 2500 counts/s while the spectrum becomes exceptionally soft. This comprehensive view of Cygnus X-1's rich and complex behavior demonstrates the clear spectral state transitions that motivate our detailed analysis of \jfw{appropriate} energy bands for characterizing these states. \jfw{The resulting hardness-intensity diagram (Section \ref{sec:hid}) confirms the expected bimodal state distribution characteristic of this system.}

\subsubsection{Orbital Phase Coverage}

Given Cygnus X-1's 5.6-day orbital period and the temporal distribution of our observations, we examined the orbital phase coverage achieved by our \textit{NuSTAR} campaign. Using the orbital ephemeris of \jfw{\citet{1999A&A...343..861B}} with $P = 5.599829$ days and $T_0 = $ MJD 41874.207, we computed the orbital phase for each observation segment.

\begin{figure*}
    \centering
    \includegraphics[width=\textwidth]{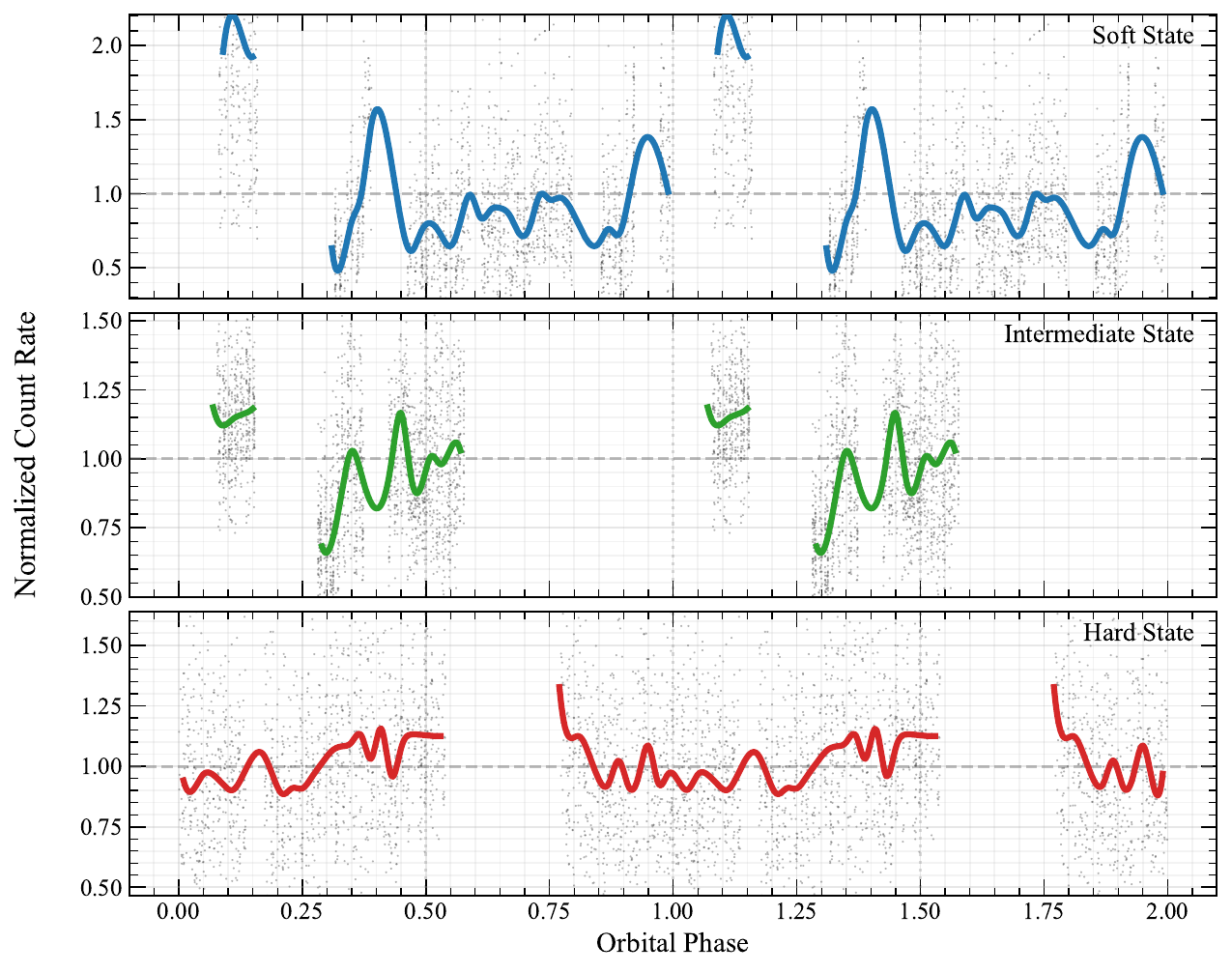}
    \caption{Orbital phase coverage of \textit{NuSTAR} observations separated by spectral state. Gray points show individual 100s time bins phase-folded using the 5.6-day orbital period, while colored lines represent smoothed profiles normalized to the mean count rate (horizontal dashed line). Data are shown over two cycles for clarity. Soft state (top): nearly complete phase coverage. Intermediate state (middle): approximately 50\% coverage concentrated at phases 0.0-0.4. Hard state (bottom): comprehensive coverage across most of the orbit.}
    \label{fig:phase_coverage}
\end{figure*}
Figure \ref{fig:phase_coverage} displays the orbital phase sampling across the three spectral states identified from our hardness ratio analysis. The soft state observations (top panel) provide nearly complete orbital coverage with enhanced sampling near phase \jfw{0.1}, fortuitously capturing the system during multiple orbital cycles. The intermediate state (middle panel) achieves approximately 50\% phase coverage, with observations concentrated between phases 0.0-0.4 \jfw{and a secondary cluster near phase 0.1.} The hard state (bottom panel) offers the most comprehensive coverage, with observations distributed across nearly the entire orbital cycle, enabling robust characterization of any orbital-dependent phenomena.

This differential phase coverage must be considered when interpreting state-dependent orbital modulation patterns. \jfw{The excellent hard and soft state sampling across nearly all orbital phases confirms that spectral state occurrence is not preferentially associated with any particular orbital phase, consistent with spectral states being driven by accretion flow properties rather than orbital geometry.}

\subsubsection{Energy Band Selection}

\begin{figure*}
    \centering
    \includegraphics[width=\textwidth]{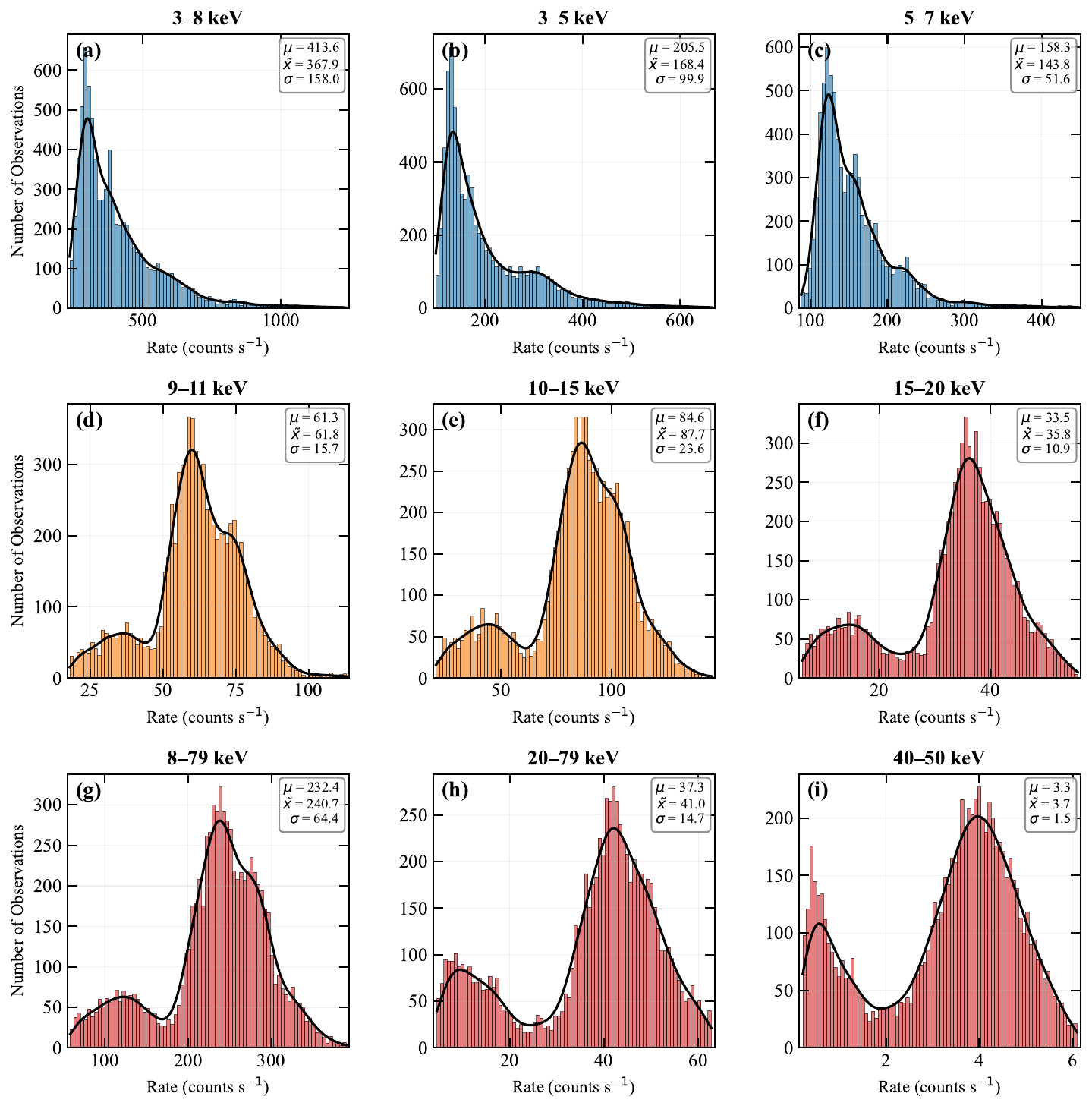}
    \caption{\jfw{Systematic energy-dependent emergence of bimodality in the count rate distributions of Cygnus X-1, constructed from 100\,s binned light curves. Panels progress from soft to hard energy bands: (a) 3--8 keV, (b) 3--5 keV, (c) 5--7 keV show unimodal distributions (blue); (d) 9--11 keV, (e) 10--15 keV, (f) 15--20 keV show the progressive emergence of a second component (orange/red); (g) 8--79 keV, (h) 20--79 keV, (i) 40--50 keV exhibit clear bimodality (red). Black curves show kernel density estimates. The bimodality is absent below $\sim$9 keV, first emerges near 10 keV, and is present in all bands above 15 keV. The broad 8--79 keV band captures the full bimodal hard component, motivating our choice of 8 keV as the soft/hard boundary.}}
    \label{fig:bimodal-appearance}
\end{figure*}

The distinct spectral states evident in Figure \ref{fig:longtermlc} require careful selection of energy bands for constructing a robust hardness ratio. \jfw{Given that specific energy ranges are highly instrument-dependent and must be tailored to both source characteristics and detector response---with different boundaries proving effective for different missions (e.g., RXTE, MAXI, Swift)---we performed a data-driven analysis to define appropriate bands for tracking Cygnus X-1's state transitions with NuSTAR.}
We \jfw{examined count rate distributions across nine energy bands spanning the full NuSTAR bandpass. Figure \ref{fig:bimodal-appearance} presents a systematic progression from soft to hard bands, revealing the energy at which the well-known bimodal state distribution emerges.}

\begin{figure*}
    \centering
    \includegraphics[scale=0.36]{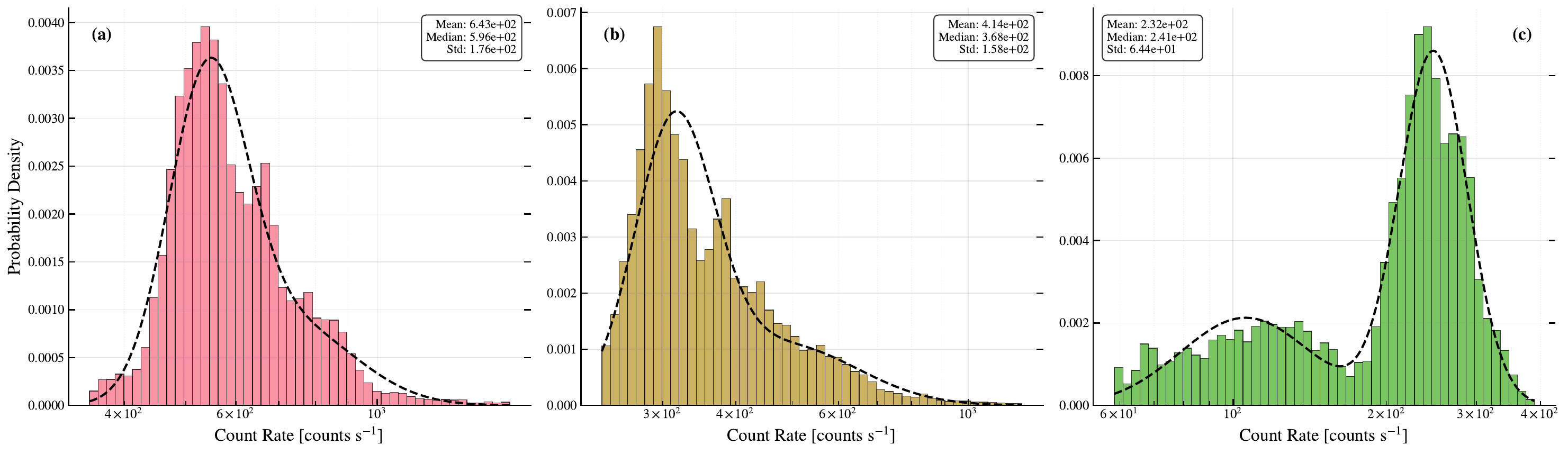}
    \caption{Distributions of the count rates for (a) the full band, (b) the soft band, and (c) the hard band. The histograms are shown on a logarithmic rate axis. The dashed line in each panel represents the best-fit model as determined by the BIC. \jfw{The full band count rate equals the sum of the soft and hard band count rates by construction, as these energy ranges are contiguous and non-overlapping. Consequently, the peak of the full band distribution ($\sim$520 counts/s) corresponds approximately to the sum of the soft ($\sim$285 counts/s) and hard ($\sim$230 counts/s) band peaks.}}
    \label{fig:distribution-fits}
\end{figure*}

\jfw{The bimodal nature of Cygnus X-1's flux distribution, reflecting its well-documented hard and soft spectral states, has been established by previous studies using long-term monitoring data \citep{2021JHEAp..31...23D, 2013A&A...554A..88G}. Importantly, \citet{10.1111/j.1365-2966.2007.12688.x} demonstrated clear bimodality in the hardness ratio (5--12/1.3--5 keV) of Cygnus X-1 using RXTE/ASM data, confirming that the source occupies two distinct spectral states. Our NuSTAR observations go further by demonstrating that the raw count rate distributions themselves become bimodal in individual energy bands above $\sim$10 keV---a distinct result showing that the flux bimodality is intrinsic to the hard X-ray spectral component, not merely a property of spectral shape ratios.} \jfw{The distributions reveal a clear energy-dependent transition. The soft bands---3--8 keV (panel a), 3--5 keV (panel b), and 5--7 keV (panel c)---all exhibit broad, unimodal distributions with no evidence of bimodality. A subtle second component begins to emerge in the 9--11 keV band (panel d), becoming progressively more distinct in the 10--15 keV (panel e) and 15--20 keV (panel f) bands. The broad 8--79 keV band (panel g), as well as the 20--79 keV (panel h) and 40--50 keV (panel i) bands, show unambiguous bimodality with well-separated peaks.}
This energy-dependent structure has clear physical origins. Below 8 keV, thermal disk emission dominates, \jfw{with disk temperature spanning a range from $\sim$0.2 keV (truncated hard state) to $\sim$0.4--0.5 keV (extended soft state)}. Above 8 keV, Comptonized emission dominates: either from an extended, geometrically thick corona (hard state) producing substantial hard X-rays, or from a largely suppressed corona (soft state) with minimal hard emission. The bistable nature of these configurations—driven by thermal and radiative instabilities—creates the observed bimodal distribution in hard X-rays while preserving continuous variation in the thermal component.
\jfw{This systematic analysis demonstrates that bimodality in narrow bands first becomes apparent around 10 keV and strengthens at higher energies. The broad 8--79 keV band captures this entire bimodal hard component, providing clear state separation. Combined with NuSTAR's peak effective area near this energy, motivated} our band definitions: 3-8 keV (soft) and 8-79 keV (hard), naturally separating thermal from Comptonized emission.
To quantify these distributions, we performed model comparison using Normal, Lognormal, and bimodal variants (Gaussian Mixture Models), selecting optimal models via the Bayesian Information Criterion. Figure \ref{fig:distribution-fits} presents the results.

While both soft and total bands are best described by overlapping bimodal lognormal models, the hard band exhibits a distinct two-component GMM with well-separated peaks. The hard band's well-separated components, in contrast to the soft band's overlapping distributions, reflect the bistable nature of the Comptonizing corona and motivate our choice of the 8--79 keV / 3--8 keV hardness ratio \jfw{for tracking Cygnus X-1's state transitions with NuSTAR}.

\textbf{Below 8 keV}: Emission is primarily from the thermal accretion disk, whose temperature and luminosity vary with radial extent. \jfw{Disk temperature values span a continuous range from $\sim 0.2$ keV \citep{2017MNRAS.472.4220B} (in the truncated hard state) to $\sim 0.4-0.5$ keV (in the extended soft state, reaching the ISCO) \citep{2016ApJ...826...87W}. The broad, unimodal distribution in this band is consistent with the source occupying a continuum of thermal states, though we note that the histogram shape alone does not constrain the temporal nature of transitions---rapid transitions through intermediate flux values cannot be excluded. This unimodal character is further understood through the finding of \citet{2013A&A...554A..88G} that the count rate--photon index relationship is non-monotonic at soft X-ray energies, making count rate alone an inherently less effective state discriminator below $\sim$8 keV.}

\textbf{Above 8 keV}: The emission is dominated by Comptonization in the corona and/or jet base. In the hard state, a geometrically thick, optically thin corona generates substantial hard X-ray emission via inverse Compton scattering. In the soft state, this corona is largely suppressed or collapsed, reducing hard X-ray production. The intermediate state reflects brief transitions between these configurations. As the corona is either extended (hard state) or suppressed (soft state), with limited time in transitional geometries, the hard X-ray flux naturally bifurcates into two distinct levels.

The increasing separation between peaks at higher energies \jfw{(see Figure \ref{fig:bimodal-appearance}, panels e--i)} further supports this interpretation. As the energy (or temperature) of the Comptonizing electrons increases, the Comptonization efficiency rises,
with the extended hard-state corona upscattering photons to higher energies than the collapsed soft-state corona. This bimodality reflects the bistable nature of accretion flows around black holes, driven by thermal and radiative instabilities that push the system toward either a cool, geometrically thin disk or a hot, geometrically thick flow, with rapid transitions between these configurations.

\jfw{Having established appropriate energy ranges for NuSTAR}, we next sought to quantitatively characterize the statistical nature of the count rate distributions within them. We analyzed the distributions for the total (3-79 keV), soft (3-8 keV), and hard (8-79 keV) bands individually. For each distribution, we performed a model comparison to determine the best statistical description. The candidate models included a standard Normal distribution, a Lognormal distribution (which often characterizes accretion processes), and their bimodal counterparts, modeled as a Gaussian Mixture Model (GMM) with two components. The optimal model for each band was selected by finding the one with the lowest Bayesian Information Criterion (BIC), which penalizes model complexity to avoid overfitting.

The results of this analysis are presented in Figure \ref{fig:distribution-fits}. The distributions for the total band (panel a) and the soft band (panel b) are both best described by an overlapping bimodal lognormal model. Crucially, while all bands exhibit bimodality, the nature of this bimodality differs significantly. In the soft band, the two lognormal components overlap substantially, creating a broad, nearly continuous distribution \jfw{without clear separation between states}. In stark contrast, the hard-band (panel c) distribution is unambiguously bimodal and is best fit by a two-component GMM with well-separated peaks. This contrast informs our choice of band ratio: the hard band's bimodality provides clean separation between states, \jfw{while the soft band's overlapping nature reflects the continuous range of thermal disk temperatures across states}.

\subsection{Fractional Variability and Modulation Factors}

The fractional variability ($F_{var}$) is a crucial diagnostic for understanding the intrinsic variability in the X-ray light curves of Cygnus X-1. It is calculated as the normalized rms fluctuation in the count rates, which allows us to distinguish between true source variability and statistical noise. We apply the method outlined by \citet{2003MNRAS.345.1271V} to obtain $F_{var}$ and its associated uncertainty.

For the majority of observations, we calculate the excess variance, and the resulting $F_{var}$ values are reported in Table \ref{tab:fracvar_results}. However, the excess variance is negative in one case (ObsID 30302019006), which suggests that the intrinsic variability of the source is too low to be detected beyond the noise level. Therefore we instead provide a 95\% confidence upper limit, derived by considering the uncertainty in the measured excess variance and applying a statistical approach to estimate the maximum possible value of $F_{var}$. This upper limit reflects the sensitivity of our measurements and indicates that the variability in this observation is consistent with no intrinsic variability, within the limits of our measurement errors.

We also calculate the modulation factor for each observation, defined as the peak-to-trough amplitude normalized by the mean count rate. The modulation factors range from 0.385 to 0.793 and show strong correlation with $F_{var}$ (Pearson $r = 0.91$), confirming that both metrics trace the same underlying variability. The highest modulation factors ($>0.75$) occur during soft states and state transitions ($F_{var} > 30\%$), while the lowest ($<0.52$) correspond to stable hard or intermediate states. The anomalous observation 30302019006 exhibits the lowest modulation factor (0.385) alongside its non-detection in $F_{var}$, reinforcing that the low variability is a genuine physical characteristic of this possible failed state transition rather than a statistical artifact.

\begin{deluxetable*}{lccccc}
\caption{Log of observations showing the start and end orbital phases and the corresponding fractional variability and modulation factors. ObsIDs marked with an asterisk (*) cross the phase 0.0 boundary. For observations where the excess variance was negative, we report the 95\% confidence upper limit on the fractional variability.}
\label{tab:fracvar_results}
\tablewidth{0pt}
\tablehead{
\colhead{OBSID} & \colhead{Start Phase} & \colhead{End Phase} & \colhead{Mean Flux (counts/s)} & \colhead{F$_{var}$ (\%)} & \colhead{Modulation Factor}
}
\startdata
30001011002 & 0.854 & 0.921 & 623.4250 & 37.30 $\pm$ 0.05 & 0.793 \\
10014001001 & 0.975 & 0.993 & 865.9116 & 29.16 $\pm$ 0.10 & 0.759 \\
30001011005 & 0.282 & 0.373 & 638.2865 & 21.79 $\pm$ 0.05 & 0.620 \\
30001011007 & 0.910 & 1.110 & 540.8489 & 14.16 $\pm$ 0.04 & 0.567 \\
30001011009 & 0.464 & 0.590 & 587.0555 & 34.52 $\pm$ 0.04 & 0.781 \\
30001011011 & 0.445 & 0.578 & 839.4739 & 12.65 $\pm$ 0.05 & 0.505 \\
30101022002 & 0.425 & 0.563 & 686.7302 & 9.84 $\pm$ 0.06 & 0.509 \\
90101020002 & 0.807 & 0.886 & 587.0142 & 16.32 $\pm$ 0.06 & 0.576 \\
30002150002 & 0.826 & 1.120 & 551.4538 & 21.59 $\pm$ 0.03 & 0.646 \\
30002150004 & 0.173 & 0.480 & 589.8527 & 20.88 $\pm$ 0.03 & 0.629 \\
30002150008 & 0.893 & 1.055 & 524.3749 & 20.92 $\pm$ 0.04 & 0.641 \\
30202032002 & 0.058 & 0.137 & 543.3733 & 16.35 $\pm$ 0.06 & 0.590 \\
30302019002 & 0.717 & 0.795 & 836.7514 & 29.00 $\pm$ 0.05 & 0.716 \\
30302019004 & 0.879 & 0.958 & 752.2041 & 16.57 $\pm$ 0.05 & 0.564 \\
30302019006 & 0.078 & 0.154 & 872.8378 & $<$1.38\% & 0.385 \\
30302019010 & 0.081 & 0.160 & 1568.9739 & 35.69 $\pm$ 0.04 & 0.757 \\
30302019012 & 0.613 & 0.702 & 766.3174 & 32.18 $\pm$ 0.05 & 0.746 \\
80502335002 & 0.945 & 1.031 & 674.9547 & 11.11 $\pm$ 0.07 & 0.519 \\
80502335006 & 0.632 & 0.711 & 479.6962 & 29.12 $\pm$ 0.06 & 0.736 \\
30702017006 & 0.773 & 0.852 & 770.8656 & 13.16 $\pm$ 0.06 & 0.526 \\
90802013002 & 0.293 & 0.372 & 711.1962 & 11.92 $\pm$ 0.06 & 0.523 \\
90802013004 & 0.461 & 0.540 & 672.0306 & 9.49 $\pm$ 0.07 & 0.508 \\
80902318002 & 0.701 & 0.792 & 656.0478 & 16.41 $\pm$ 0.05 & 0.581 \\
80902318004 & 0.316 & 0.390 & 724.7160 & 40.08 $\pm$ 0.05 & 0.787 \\
30901039002 & 0.843 & 0.933 & 500.8741 & 17.74 $\pm$ 0.06 & 0.633 \\
91002320004 & 0.784 & 0.862 & 547.2458 & 16.51 $\pm$ 0.06 & 0.605 \\
\enddata
\end{deluxetable*}

The non-detection of variability in observation 30302019006, despite the source exhibiting a clear intermediate state, is striking. This state is canonically associated with high levels of aperiodic variability as the accretion flow undergoes significant reconfiguration \citep{10.1093/mnras/stx2110}. The observed quiescence suggests a decoupling of the spectral and timing evolution. One possible interpretation is that this observation captured a so-called `failed` or `aborted` state transition (for a more detailed analysis, see \cite{2008mqw..confE..89W}). The accretion disk may have begun to cool and move inwards, softening the spectrum, but the physical mechanism responsible for producing strong variability (e.g., instability at the disk-corona interface \citep{2001MNRAS.321..759C}) failed to fully develop. This anomalously stable intermediate state highlights the complex nature of state transitions and demonstrates that the connection between the accretion geometry and its variability signature is not always straightforward.

\begin{figure*}
    \centering
    \includegraphics[scale=0.7]{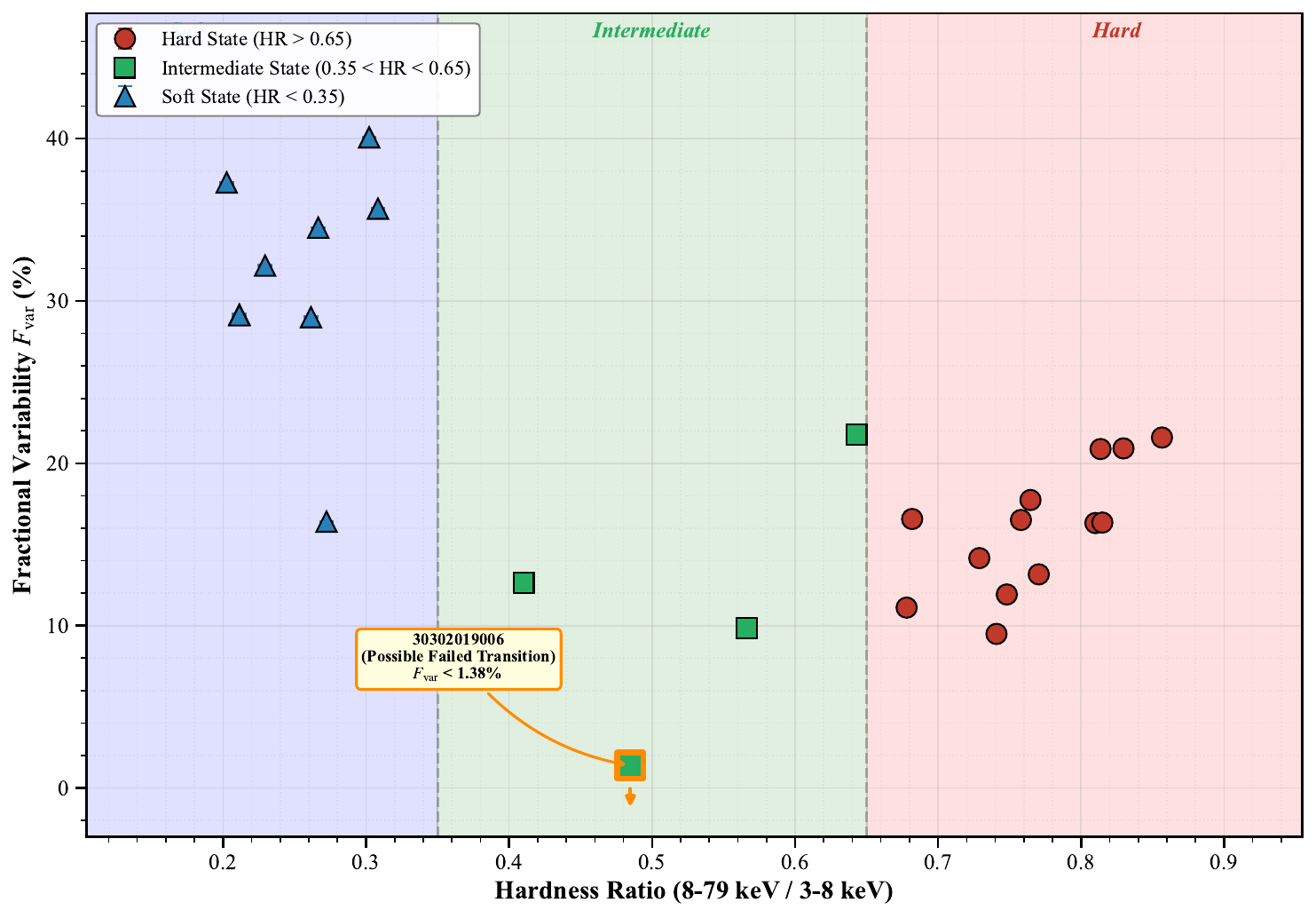}
    \caption{Fractional variability ($F_{var}$) versus hardness ratio for all 26 NuSTAR observations. Blue triangles represent soft state observations (HR < 0.35), green squares indicate intermediate state (0.35 < HR < 0.65), and red circles denote hard state (HR > 0.65). Background shading delineates the spectral state regions. The anomalous observation 30302019006 is highlighted with an orange annotation, showing its upper limit of $F_{var} < 1.38\%$ despite intermediate hardness---evidence of a \jfw{likely} failed state transition where spectral softening occurred without the accompanying increase in variability.}
    \label{fig:fvar_hardness}
\end{figure*}

\jfw{Figure \ref{fig:fvar_hardness} displays the relationship between fractional variability and spectral hardness across all observations. The soft state observations (HR < 0.35) exhibit the highest variability, with $F_{var}$ values ranging from 29\% to 40\%, while hard state observations (HR > 0.65) show more moderate variability between 10\% and 22\%. The anomalous observation 30302019006 stands out dramatically in the intermediate state region, with its upper limit of $F_{var} < 1.38\%$ lying far below all other observations regardless of spectral state. This figure provides the clearest visualization of the failed state transition: despite having an intermediate hardness ratio (HR $\approx$ 0.49), this observation exhibits variability suppressed by more than an order of magnitude compared to other intermediate state observations.}

\subsection{The Hardness-Intensity and Color-Color Diagrams}
\label{sec:hid}
\begin{figure*}
    \centering
    \includegraphics[scale=0.45]{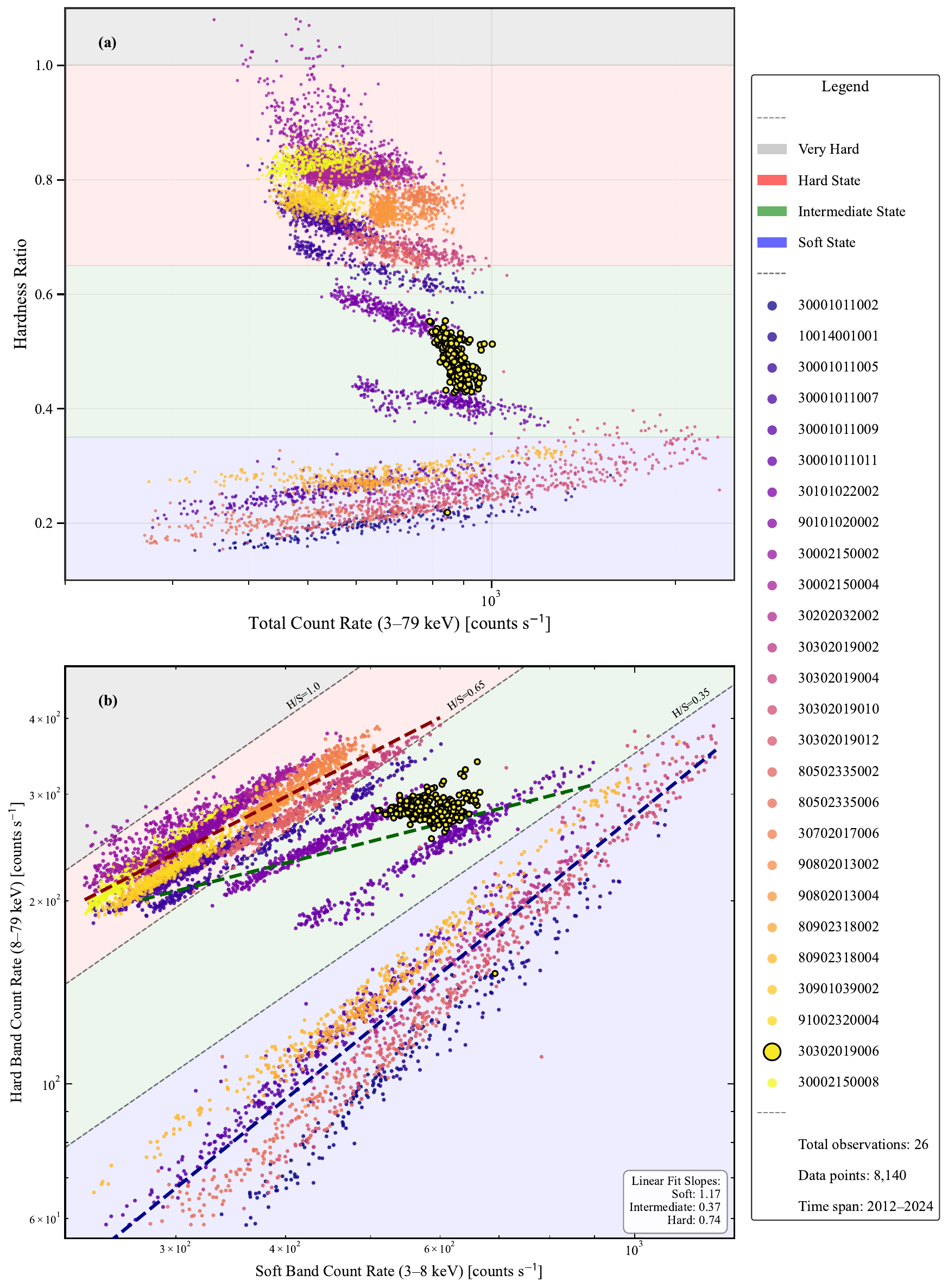}
    \caption{X-ray spectral state classification from 2012--2024 \textit{NuSTAR} observations. (a) Hardness-intensity diagram with background shading indicating spectral state regions: gray (very hard), red (hard), green (intermediate), and blue (soft). (b) Hard versus soft band count rate correlation, with diagonal dashed lines at constant hardness ratios of 1.0, 0.65, and 0.35. Both panels show 8,140 time bins from 26 epochs, color-coded chronologically. The bimodal distribution separates the hard state (lower branch) from the soft state (upper branch). Observation 30302019006 is highlighted for its anomalous anticorrelation between bands.}
    \label{fig:HID}
\end{figure*}

To more clearly delineate the relationship between the spectral states, we constructed a hardness-intensity diagram (HID) for the entire dataset, shown in Figure \ref{fig:HID}. Each point corresponds to a 100 s time bin, with its position determined by the total 3-79 keV count rate (Intensity) and the hardness ratio (HR). The data points are color-coded by observation, illustrating how the source explores the diagram over the full 12-year observing baseline.

The hardness-intensity diagram (HID) and flux-flux correlation plots provide fundamental diagnostics for classifying the accretion states of Cygnus X-1. Unlike many transient X-ray binaries that exhibit clear q-shaped hysteresis loops \citep{2025ApJ...985..217H}, Cygnus X-1 displays a more persistent behavior \citep{2016ASSL..440...61B}, spending approximately 50\% of its time in the hard state with occasional transitions to softer states.
The HID analysis reveals two distinct branches in the intensity-hardness plane (see also \cite{2020A&A...637A..66M} and \cite{2011A&A...533A...8B}). \jfw{The HID reveals the well-known bimodal distribution characteristic of Cygnus X-1. The hard state (HR > 0.65) exhibits count rates spanning approximately 400-900 counts/s with a peak around 550 counts/s, while the soft state (HR < 0.35) shows higher count rates reaching up to 1600 counts/s. The intermediate state (0.35 < HR < 0.65) bridges these regimes across a broad range of intensities.} \jfw{Hard state observations generally exhibit a mild anticorrelation between intensity and hardness in the HID. However, observation 30302019006 displays a notably steeper anticorrelation slope, with its anomalous behavior more clearly evident in the flux-flux diagram (Figure \ref{fig:HID}b), where it shows a distinct negative correlation between soft and hard band fluxes.} We interpret this as a signature of failed state transitions commonly observed in wind-fed systems.
The flux-flux diagram provides complementary state classification through the correlation between soft (3-8 keV) and hard (8-79 keV) band count rates. The data follow distinct tracks corresponding to different hardness ratios, with the hard state observations clustering along steeper slopes (H/S > 0.65) and soft state observations following shallower trajectories (H/S < 0.35). The power-law correlation between bands remains approximately linear in log-log space within each state, with correlation indices ranging from 1.2-1.5 for the hard state to 0.8-1.0 for the soft state.
The absence of a complete hysteresis loop and the dominance of hard state observations (comprising $\sim13$ of the 26 observation epochs) confirm that Cygnus X-1 remains primarily in a low/hard state configuration. State transitions occur on timescales of weeks to months rather than the rapid cycling observed in more active systems, consistent with the system's persistent nature and its relatively stable mass accretion rate of approximately 0.01-0.1 $L_{Edd}$ \citep{10.1093/mnras/stt1044}.

\begin{figure*}
    \centering
    \includegraphics[scale=0.58]{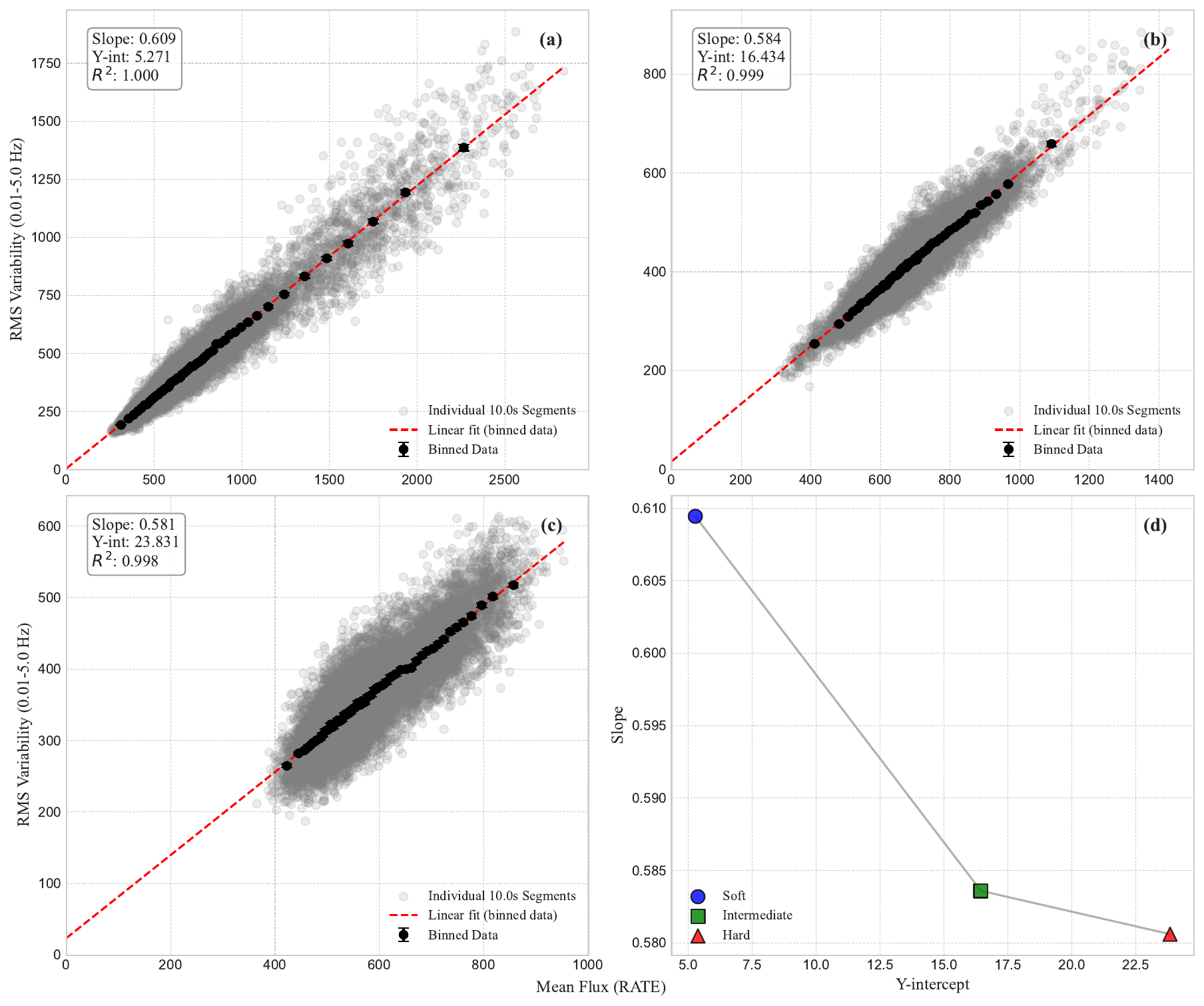}
    \caption{Linear rms-flux correlations across spectral states. (a--c) RMS variability (0.01--5.0 Hz) versus mean count rate for the soft (a), intermediate (b), and hard (c) states. Gray points show individual 10.0\,s segments; black points with error bars show binned averages; red dashed lines show best-fit linear relations. Slopes decrease from 0.609 (soft) to 0.584 (intermediate) to 0.581 (hard), while y-intercepts increase from 5.3 to 16.4 to 23.8, reflecting the growing contribution of intrinsically variable non-thermal emission. (d) Parameter evolution summary. The linear rms-flux scaling across all states confirms multiplicative coupling between accretion fluctuations and radiative output, modulated by state-dependent geometry.}
    \label{fig:rms_flux}
\end{figure*}

\jfw{A fundamental prediction of the propagating fluctuations model is that the rms variability amplitude should scale linearly with mean flux \citep{2001MNRAS.323L..26U}. To test this relationship across spectral states, we computed the rms variability in the 0.01--5.0 Hz frequency range for 10-second segments throughout each observation. Figure \ref{fig:rms_flux} presents the resulting rms-flux relations separated by spectral state. All three states exhibit clear linear correlations, confirming that the multiplicative coupling between accretion rate fluctuations and radiative output operates universally in Cygnus X-1. However, the parameters of these relations evolve systematically with spectral state. The soft state shows the steepest slope (0.609) with the smallest y-intercept (5.3), indicating that variability scales nearly proportionally with flux when thermal disk emission dominates. The intermediate and hard states display progressively shallower slopes (0.584 and 0.581) with increasing y-intercepts (16.4 and 23.8), reflecting the growing contribution of intrinsically variable non-thermal emission from the corona and jet base. This systematic evolution of rms-flux parameters provides independent confirmation of the changing relative contributions of disk and corona across spectral states inferred from the HID analysis.}

\subsection{Power Spectral Density Analysis}

We performed power spectral density (PSD) analysis on all 26 \textit{NuSTAR} observations to characterize the stochastic variability properties across different accretion states. Our approach is complementary to recent comprehensive spectral-timing studies of Cygnus X-1 performed with NICER, which have focused on the rich variability present at lower energies (< 2 keV) where the accretion disk emission dominates \citep{2024A&A...687A.147K}. The PSDs were computed using Welch's method \citep{1161901} with 256-second segments and 50\% overlap, providing frequency coverage from 0.01 to 5 Hz. Our frequency range is bounded by the Nyquist limit at high frequencies, \jfw{while the lower limit of 0.01 Hz is determined by both the segment duration (256 s, corresponding to a minimum resolvable frequency of $\sim$0.004 Hz) and the dominance of instrument noise at lower frequencies.} To emphasize characteristic timescales while preserving spectral shape information, we present the PSDs in frequency-weighted form ($f \times \text{PSD}$ vs. $f$), where peaks directly indicate the frequencies containing maximum variability power.

This frequency band captures an intermediate portion of the well-documented broadband noise spectrum of Cygnus X-1, which is known to span from approximately $10^{-3}$ Hz to $\sim$100 Hz. The broadband power in the hard and intermediate states is typically modeled as a superposition of multiple Lorentzian components \citep{2003A&A...407.1039P}. In our observed band, a single dominant component is evident below 1 Hz. While the onset of a higher-frequency component is visible above this, its characteristic frequency lies beyond our reliable range of analysis. Consequently, to avoid biases from fitting an incomplete feature, we modeled the spectrum exclusively within the 0.01–1 Hz range, employing a single broad Lorentzian function to accurately characterize the dominant, fully-captured noise process. 

\begin{figure*}
    \centering
    \includegraphics[scale=0.54]{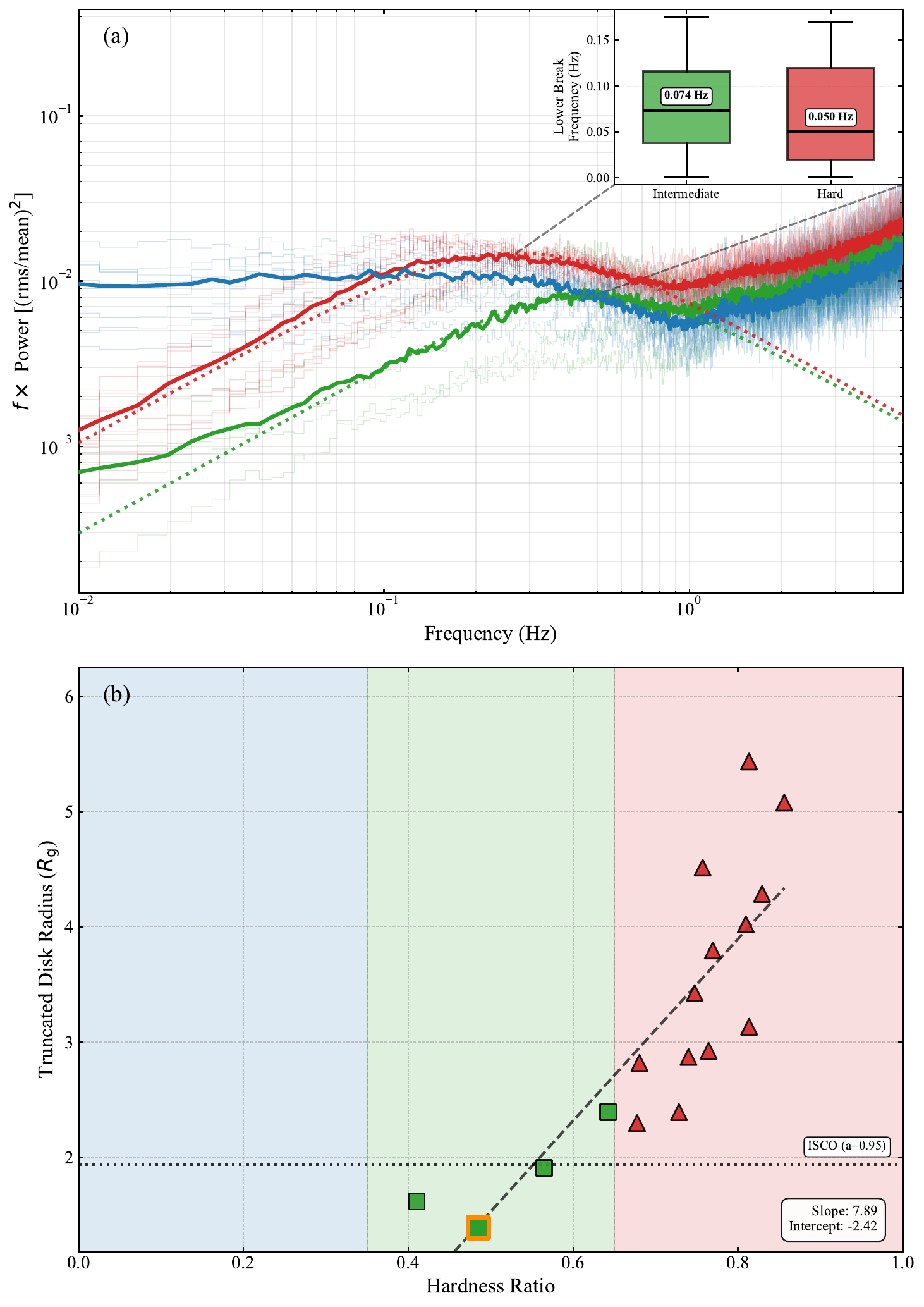}
    \caption{Frequency-weighted PSD analysis. (a) $f \times \text{PSD}$ curves color-coded by spectral state; semi-transparent lines show individual observations, bold lines show state averages. Hard and intermediate states show break frequencies \jfw{with medians of $\sim$0.050 Hz and $\sim$0.074 Hz (inset),} plus a feature near $\sim$1 Hz absent in the featureless soft state. (b) Truncation radius versus hardness ratio from a truncated disk model. The dashed line shows the linear fit; the dotted line marks the ISCO for $a = 0.95$. Observation 30302019006 (orange) shows anomalously small truncation ($<1R_g$) despite intermediate hardness.}
    \label{fig:PSD_analysis}
\end{figure*}

Figure \ref{fig:PSD_analysis} displays the frequency-weighted Power Spectral Densities (PSDs), color-coded by accretion state. While individual PSDs (semi-transparent) show significant scatter, the mean trends (bold lines) reveal a clear evolution. The characteristic break frequency, seen as a peak in this representation, systematically shifts to higher values as the source softens. In the hard state, the break is narrowly distributed around a median of 0.050 Hz (see \ref{fig:PSD_analysis}a inset). This shifts to a higher and more broadly distributed median of 0.074 Hz in the intermediate state. In contrast, the soft state exhibits a featureless red-noise spectrum with no discernible break. This frequency increase suggests a decrease in the characteristic variability timescale, consistent with a shrinking Comptonizing region as the inner disk moves closer to the black hole.

We assume a truncated geometrically thin-disk model and convert the characteristic frequencies to physical scales using the empirical relation $R_g \sim 6 \left( \frac{50}{m \cdot f_{LB}} \right)^{\frac{2}{3}}$ \citep{2007A&ARv..15....1D} for $m = 21.2$ M$_{\odot}$, where $R_g$ is the truncation radius and $f_{LB}$ is the lower break frequency. We observe a roughly linear trend between $f_{LB}$ and $R_g$, with ranging from $\sim$2 $R_g$ in intermediate states to $\sim$5.5 $R_g$ in hard states (Figure \ref{fig:PSD_analysis}b).

Notably, observation 30302019006 (highlighted with orange edge in Figure \ref{fig:PSD_analysis}b) exhibits anomalous behavior despite its intermediate spectral hardness. This observation also displayed anticorrelated hard and soft band fluxes in the color-color diagram (Figure \ref{fig:HID}b), where typically these bands show positive correlation. We interpret this as a failed state transition: the disk approached the ISCO (evidenced by the high break frequency) but failed to establish the stable soft-state configuration, possibly due to insufficient accretion rate or disruption by the stellar wind. The absence of quasi-periodic oscillations (QPOs) in any observation above 3$\sigma$ significance is consistent with Cygnus X-1's behavior outside rare transitional episodes \citep{2025A&A...696A.237F}. The systematic evolution of break frequencies and truncation radii demonstrates that geometric reconfiguration of the accretion flow drives both spectral and timing properties in this archetypal black hole system.

\subsubsection{Frequency-Dependent Time Lag Analysis}

To probe the causal connections between soft and hard X-ray emission, we computed frequency-dependent time lags between the soft and hard bands using cross-spectral analysis. Time lags encode crucial information about the propagation of accretion rate fluctuations and reprocessing geometry in the inner accretion flow. Previous studies \citep{2001ChA&A..25..416Q, 2000A&A...357L..17P} have revealed that hard X-rays consistently lag behind soft X-rays in all states, with lag values depending on frequency and energy separation. During state transitions, the X-ray lag exhibits significant variability in both its shape and magnitude, often reaching amplitudes substantially larger than those observed during periods of spectral stability.

We calculated cross-power spectra using Welch's method with adaptive window sizes (256-1024 points) to optimize frequency resolution while maintaining adequate statistics. The time lag at each frequency was derived from the phase of the cross-spectrum: $\tau(\nu) = -\phi_{xy}(\nu)/(2\pi\nu)$, \jfw{where $\phi_{xy}(\nu) = \arg[C_{xy}(\nu)]$ is the phase angle of the cross-power spectrum between the soft (x) and hard (y) bands, and} positive lags indicate hard photons lagging soft photons. To ensure reliability, we required coherence $>0.2$ and applied 5$\sigma$ outlier rejection to the lag measurements. The lags were computed in 40 logarithmically-spaced frequency bins spanning 0.01-5 Hz.

\begin{figure*}
    \centering
    \includegraphics[width=\textwidth]{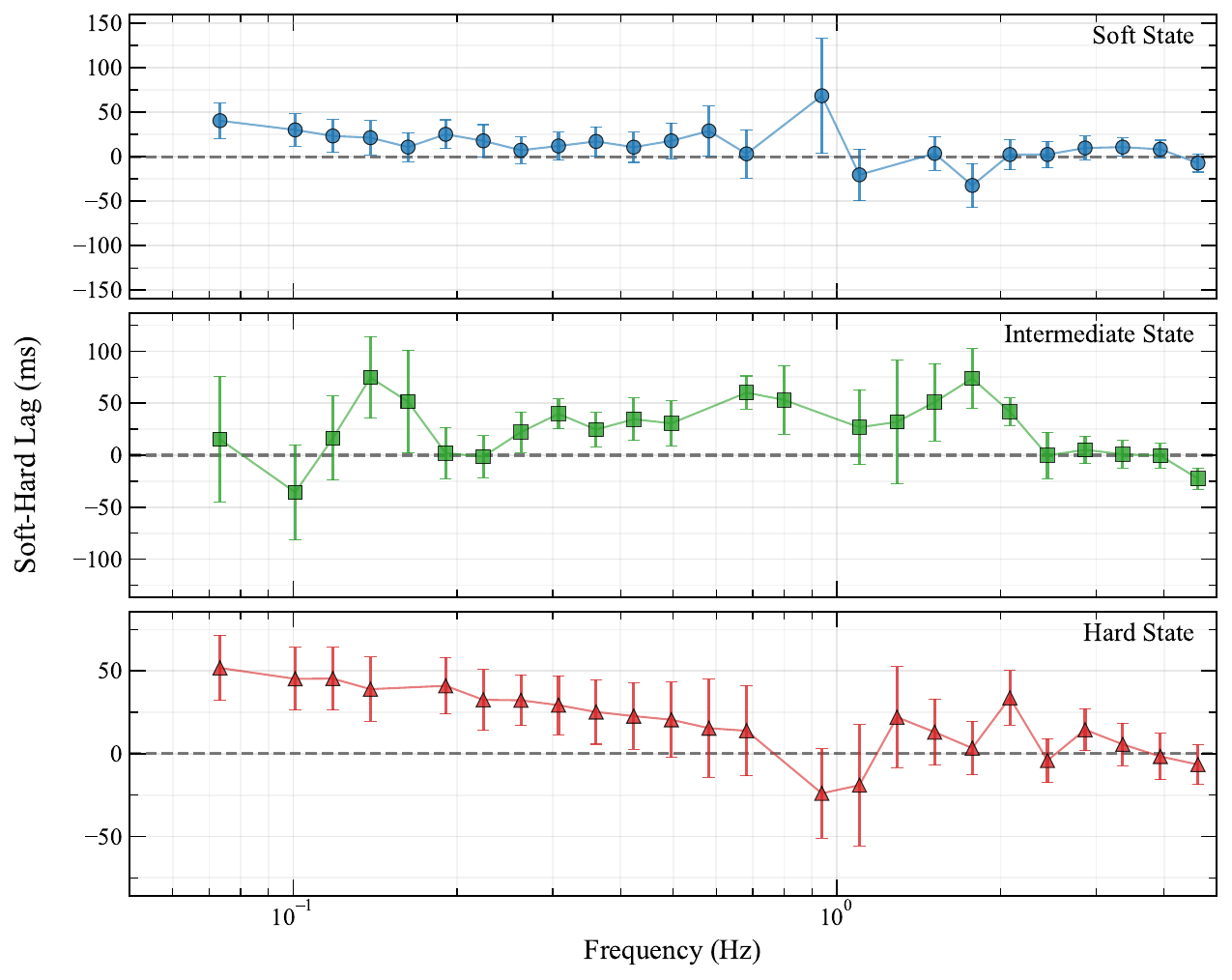}
    \caption{Frequency-dependent soft-hard time lags across spectral states. Top: Soft state shows near-zero lags, consistent with minimal Comptonization. Middle: Intermediate state exhibits complex behavior transitioning from near-zero to positive (hard-leading) lags around 0.3 Hz. Bottom: Hard state displays systematic positive lags decreasing from $\sim$45 to $\sim$20 ms, characteristic of Comptonization delays. Error bars represent 1$\sigma$ uncertainties from \jfw{coherence-weighted cross-spectral analysis}. The lag evolution directly probes the geometric transformation between accretion configurations.}
    \label{fig:lag_frequency}
\end{figure*}
Figure \ref{fig:lag_frequency} shows the resulting lag-frequency spectra. \jfw{For each spectral state, we computed time lags from all continuous data segments within observations classified in that state, then calculated coherence-weighted averages across segments within each frequency bin.} The results reveal a clear and systematic evolution of the time lags as Cygnus X-1 transitions between states.

In the soft state (top panel), the low-frequency hard lag is strongly suppressed, remaining close to zero below 0.5 Hz. This quenching of the lag is a key indicator of a fundamental change in the accretion geometry. It is consistent with the inner edge of the accretion disk extending much closer to the black hole, drastically reducing the size of the hard X-ray emitting corona and, consequently, the light-travel times associated with propagating fluctuations.

As the source transitions to the intermediate state (middle panel), the lag spectrum evolves significantly. The lag remains hard, but its magnitude increases, reaching a maximum of approximately 75 ms near 0.2 Hz before declining towards higher frequencies. This enhancement and shift in the peak lag frequency likely correspond to changes in the structure and dominant timescale of the corona during the state transition. This state also shows the largest variability across frequency ranges, which is possibly indicative of radiation-emitting outflows in the hard state.

Finally, in the hard state (bottom panel), we observe a prominent hard lag (positive values) of approximately 50 ms at the lowest frequencies. This lag systematically decreases with increasing frequency, becoming consistent with zero above $\sim$1 Hz. This behavior is the classic signature of propagating accretion fluctuations, where variations originating in the outer regions of the Comptonizing medium travel inwards, modulating the inner regions and imprinting a characteristic frequency-dependent time delay.

At higher frequencies (near 1 Hz), the lag behavior becomes more complex, with both the hard and soft states showing features that may include a brief turn to a soft lag (negative values). Such features could be indicative of a Compton reverberation signal, where hard coronal photons are rapidly reprocessed by the optically thick accretion disk. Overall, the systematic evolution of the lag spectrum presented here provides strong evidence for the truncated disk/evolving corona model for black hole state transitions.

\section{Conclusion}

This comprehensive NuSTAR timing study establishes Cygnus X-1 as the benchmark system for understanding accretion physics in stellar-mass black holes. \jfw{Our analysis of 26 observations over 12 years reveals several relationships between spectral and timing properties that inform models of disk-corona coupling and state transitions.}
\jfw{Our systematic narrow-band decomposition reveals that raw count rate bimodality is absent below $\sim$9 keV and progressively strengthens above 10 keV, localizing the bimodal flux signature to the Comptonized spectral component.} The energy dependence of the peak separation directly reflects the distinct emission mechanisms: continuous thermal disk temperatures below 8 keV versus bistable coronal configurations above.
The systematic evolution of characteristic frequencies from 0.050 Hz to 0.074 Hz between hard and intermediate states directly maps the inward motion of the disk truncation radius. Combined with the frequency-dependent time lag evolution—from prominent 50 ms hard lags to near-zero lags—these measurements provide the most precise observational constraints to date on the changing extent of the Comptonizing region. The persistence of linear rms-flux relations across all states, with systematically evolving parameters, reveals that multiplicative coupling between accretion fluctuations and radiative output operates universally but is modulated by the state-dependent geometry.
We also identify a possible failed state transition in observation 30302019006. This observation's unique characteristics—anticorrelated band fluxes, anomalously suppressed variability despite intermediate hardness, and an apparent sub-ISCO truncation radius—cannot be explained by standard transition models. We propose this represents a new transitional pathway where the disk approaches the black hole but fails to establish a stable configuration, possibly due to disruption by the companion's stellar wind. This finding opens new avenues for theoretical investigation and suggests that state transitions in wind-fed systems may follow fundamentally different evolutionary paths than those in Roche-lobe overflow systems.
Our results have broader implications for understanding accretion physics across mass scales. The tight correlations between spectral hardness, characteristic timescales, and variability amplitudes suggest universal scaling relations that may extend from stellar-mass to supermassive black holes. The failed transition phenomenon may explain similar anomalous behavior observed in other systems, including certain changing-look AGN.

\begin{acknowledgments}
This work was partially supported by
a program of the Polish Ministry of Science under the title
“Regional Excellence Initiative,” project No. RID/SP/0050/
2024/1.
We thank the \textit{NuSTAR} team for providing high-quality X-ray observational data. \jfw{This research has made use of data and/or software provided by the High Energy Astrophysics Science Archive Research Center (HEASARC), which is a service of the Astrophysics Science Division at NASA/GSFC.} We also thank Scott Lucchini and Khalid Mohamed for their assistance with figure formatting and insightful suggestions.
\end{acknowledgments}



%
\facilities{\textit{NuSTAR}}



\bibliography{cygx}{}
\bibliographystyle{aasjournalv7}



\end{document}